\documentclass[12pt]{cernart} 
\usepackage{graphicx}

\begin{document}
\newcommand{\be}{\begin{equation}}
\newcommand{\ee}{\end{equation}} 
\newcommand{\qx }{$q(x)$}
\newcommand{\Deqx }{$\Delta q(x)$}
\newcommand{\Deqtx }{$\Delta_T q(x)$}
\newcommand{\gevc }{GeV/$c$}
\newcommand{\gevcc }{GeV/$c^2$}
\renewcommand{\textfraction}{0.01}
\renewcommand{\floatpagefraction}{0.99}

\dimen\footins=\textheight

\begin{titlepage}
\docnum{CERN--PH--EP/2010--013}
\date{18 May, 2010}
\vspace{1cm}

\title{Measurement of the Collins and Sivers asymmetries on
transversely polarised protons}

\vspace*{0.5cm}
\collaboration{The COMPASS Collaboration}

\vspace{2cm}
\begin{abstract}
The Collins and Sivers asymmetries for charged hadrons produced in deeply 
inelastic scattering on  transversely polarised protons 
have been extracted from the data collected
in 2007 with the CERN SPS muon beam tuned at 160~\gevc.
At large values of the Bjorken $x$ variable 
non-zero Collins asymmetries are observed
 both for positive and negative hadrons while the Sivers
asymmetry for positive hadrons is slightly positive over almost all
the measured $x$  range. 
These results nicely support the present
theoretical interpretation of these asymmetries, in terms of leading-twist
quark distribution and fragmentation functions.
\end{abstract}

%

\vfill
\submitted{submitted to Physics Letters B}

\begin{Authlist}
{\large  The COMPASS Collaboration}\\[\baselineskip]
{\small
%
%
M.G.~Alekseev\Iref{turin_i},
V.Yu.~Alexakhin\Iref{dubna},
Yu.~Alexandrov\Iref{moscowlpi},
G.D.~Alexeev\Iref{dubna},
A.~Amoroso\Iref{turin_u},
A.~Austregesilo\IIref{cern}{munichtu},
B.~Bade{\l}ek\Iref{warsaw},
F.~Balestra\Iref{turin_u},
J.~Ball\Iref{saclay},
J.~Barth\Iref{bonnpi}
% ...
}
\end{Authlist}
%
%
\Instfoot{bielefeld}{Universit\"at Bielefeld, Fakult\"at f\"ur Physik, 33501 Bielefeld, Germany\Aref{f}}
\Instfoot{bochum}{Universit\"at Bochum, Institut f\"ur Experimentalphysik, 44780 Bochum, Germany\Aref{f}}
\Instfoot{bonniskp}{Universit\"at Bonn, Helmholtz-Institut f\"ur  Strahlen- und Kernphysik, 53115 Bonn, Germany\Aref{f}}
\Instfoot{bonnpi}{Universit\"at Bonn, Physikalisches Institut, 53115 Bonn, Germany\Aref{f}}
\Instfoot{brno}{Institute of Scientific Instruments, AS CR, 61264 Brno, Czech Republic\Aref{g}}
\Instfoot{calcutta}{Matrivani Institute of Experimental Research \& Education, Calcutta-700 030, India\Aref{h}}
\Instfoot{dubna}{Joint Institute for Nuclear Research, 141980 Dubna, Moscow region, Russia\Aref{i}}
\Instfoot{erlangen}{Universit\"at Erlangen--N\"urnberg, Physikalisches Institut, 91054 Erlangen, Germany\Aref{f}}
\Instfoot{freiburg}{Universit\"at Freiburg, Physikalisches Institut, 79104 Freiburg, Germany\Aref{f}}
\Instfoot{cern}{CERN, 1211 Geneva 23, Switzerland}
\Instfoot{liberec}{Technical University in Liberec, 46117 Liberec, Czech Republic\Aref{g}}
\Instfoot{lisbon}{LIP, 1000-149 Lisbon, Portugal\Aref{j}}
\Instfoot{mainz}{Universit\"at Mainz, Institut f\"ur Kernphysik, 55099 Mainz, Germany\Aref{f}}
\Instfoot{miyazaki}{University of Miyazaki, Miyazaki 889-2192, Japan\Aref{k}}
\Instfoot{moscowlpi}{Lebedev Physical Institute, 119991 Moscow, Russia}
\Instfoot{munichlmu}{Ludwig-Maximilians-Universit\"at M\"unchen, Department f\"ur Physik, 80799 Munich, Germany\AAref{f}{l}}
\Instfoot{munichtu}{Technische Universit\"at M\"unchen, Physik Department, 85748 Garching, Germany\AAref{f}{l}}
\Instfoot{nagoya}{Nagoya University, 464 Nagoya, Japan\Aref{k}}
\Instfoot{praguecu}{Charles University in Prague, Faculty of Mathematics and Physics, 18000 Prague, Czech Republic\Aref{g}}
\Instfoot{praguectu}{Czech Technical University in Prague, 16636 Prague, Czech Republic\Aref{g}}
\Instfoot{protvino}{State Research Center of the Russian Federation, Institute for High Energy Physics, 142281 Protvino, Russia}
\Instfoot{saclay}{CEA IRFU/SPhN Saclay, 91191 Gif-sur-Yvette, France}
\Instfoot{telaviv}{Tel Aviv University, School of Physics and Astronomy, 69978 Tel Aviv, Israel\Aref{m}}
\Instfoot{triest_i}{Trieste Section of INFN, 34127 Trieste, Italy}
\Instfoot{triest}{University of Trieste, Department of Physics and Trieste Section of INFN, 34127 Trieste, Italy}
\Instfoot{triestictp}{Abdus Salam ICTP and Trieste Section of INFN, 34127 Trieste, Italy}
\Instfoot{turin_u}{University of Turin, Department of Physics and Torino Section of INFN, 10125 Turin, Italy}
\Instfoot{turin_i}{Torino Section of INFN, 10125 Turin, Italy}
\Instfoot{turin_p}{University of Eastern Piedmont, 1500 Alessandria,  and Torino Section of INFN, 10125 Turin, Italy}
\Instfoot{warsaw}{So{\l}tan Institute for Nuclear Studies and University of Warsaw, 00-681 Warsaw, Poland\Aref{n} }
\Instfoot{warsawtu}{Warsaw University of Technology, Institute of Radioelectronics, 00-665 Warsaw, Poland\Aref{o} }
\Instfoot{yamagata}{Yamagata University, Yamagata, 992-8510 Japan\Aref{k} }
%
%
\Anotfoot{+}{Deceased}
\Anotfoot{a}{Also at IST, Universidade T\'ecnica de Lisboa, Lisbon, Portugal}
\Anotfoot{b}{Also at Chubu University, Kasugai, Aichi, 487-8501 Japan$^{\rm j)}$}
\Anotfoot{c}{Also at KEK, 1-1 Oho, Tsukuba, Ibaraki, 305-0801 Japan}
\Anotfoot{d}{On leave of absence from JINR Dubna}
\Anotfoot{e}{Also at GSI mbH, Planckstr.\ 1, D-64291 Darmstadt, Germany}
\Anotfoot{f}{Supported by the German Bundesministerium f\"ur Bildung und Forschung}
\Anotfoot{g}{Suppported by Czech Republic MEYS grants ME492 and LA242}
\Anotfoot{h}{Supported by SAIL (CSR), Govt.\ of India}
%
%
\Anotfoot{i}{Supported by Supported by CERN-RFBR grant 08-02-91009}

\Anotfoot{j}{Supported by the Portuguese FCT - Funda\c{c}\~{a}o para a
             Ci\^{e}ncia e Tecnologia grants POCTI/FNU/49501/2002 and POCTI/FNU/50192/2003}
\Anotfoot{k}{Supported by the MEXT and the JSPS under the Grants No.18002006, No.20540299 and No.18540281; Daiko Foundation and Yamada Foundation}
\Anotfoot{l}{Supported by the DFG cluster of excellence `Origin and Structure of the Universe' (www.universe-cluster.de)}
\Anotfoot{m}{Supported by the Israel Science Foundation, founded by the Israel Academy of Sciences and Humanities}
\Anotfoot{n}{Supported by Ministry of Science and Higher Education grant 41/N-CERN/2007/0}
\Anotfoot{o}{Supported by KBN grant nr 134/E-365/SPUB-M/CERN/P-03/DZ299/2000}
~
\end{titlepage}



After first indications of transverse spin effects 
in hadron physics in the 1970s \cite{Bunc76,Antille:1980th} 
their importance was unambiguously established by the remarkably large 
single spin asymmetries (SSAs) found in $pp$ collisions at Fermilab 
both for neutral and charged pions~\cite{E704a}. 
Following the discovery by the EMC at CERN in 1988 that the quark 
spins contribute only little to the proton spin \cite{EMC88}, the interest 
in the nucleon spin structure was revived and a more complete
description including quark transverse spin and transverse momentum
has been worked out.

The quark structure of the nucleon in the collinear approximation 
or after integration over the intrinsic quark
transverse momentum $\vec k_T$ is fully specified at the twist-two
level by three parton distribution functions (PDFs)
for each quark flavour \cite{Jaffe:1991kp}: 
the momentum distributions \qx,
the helicity distributions \Deqx\ and 
the transverse spin distributions \Deqtx, where $x$ is the Bjorken variable. 
The latter distribution---often referred to as transversity---is 
chiral-odd and thus not directly observable in deep inelastic scattering (DIS).
In 1993 it was suggested \cite{Collins:1993kk} 
that transversity could be measured 
in semi-inclusive lepton--nucleon scattering (SIDIS) due to a 
mechanism involving another chiral-odd function in the hadronisation, 
known today as the Collins fragmentation function (FF). 
The mechanism leads to a left-right asymmetry in the distribution
of the hadrons produced in the fragmentation of transversely
polarised  quarks.
Thus a transverse spin dependence  in the azimuthal distributions
of the final state hadrons can be generated both  in transversely 
polarised $pp$ scattering and in SIDIS off transversely polarised nucleons.
In the latter case the measurable Collins asymmetry, $A_{Coll}$, 
is proportional to the convolution of the transversity PDF and the 
Collins FF.

Admitting a finite $\vec k_T$, in total eight PDFs are needed for
a full description at leading twist and leading order in $\alpha_S$
\cite{Kotzinian:1994dv,Mulders:1995dh,Bacchetta:2006tn}. 
All these functions lead to azimuthal 
asymmetries in the distribution of  hadrons produced  in
SIDIS processes and can be disentangled
measuring the different angular modulations.
Amongst the transverse momentum dependent PDFs,
the $T$-odd Sivers function \cite{Sivers:1989cc} is of particular interest.
This function arises from a correlation between the transverse momentum 
of an unpolarised quark in a transversely polarised nucleon and the 
nucleon spin.
It can be different from zero because of final state 
interactions mediated by soft gluon exchange between the interacting quark
and the target remnants \cite{Brodsky:2006hj}.
It is responsible for the  Sivers asymmetry, $A_{Siv}$, which
is proportional to the convolution of the Sivers function
and the unpolarised FF. 
The Sivers mechanism might also be the reason for the large 
asymmetries observed in $pp$ collisions.

Transverse spin effects in SIDIS are investigated, at different 
beam energies, by the HERMES experiment at DESY and the 
COMPASS experiment at CERN.
An experiment to measure transversity using a transversely polarised 
$^3$He target has recently been performed at JLab \cite{jlab}.
Transverse spin effects are also an important part of the scientific programme 
of the RHIC spin experiments at BNL.

Up to now, sizable
Collins asymmetries for the proton were observed recently by 
HERMES using a proton target \cite{Airapetian:2004tw}.
This implies non-vanishing Collins fragmentation and transversity functions.
Direct measurements at the KEK $e^+e^-$ collider by the BELLE experiment  
established  that this Collins FF is 
sizable \cite{Abe:2005zx,Seidl:2008xc}.
COMPASS  measured vanishing asymmetries by scattering high energy muons off 
a deuteron target~\cite{Alexakhin:2005iw,Ageev:2006da,Alekseev:2008dn}.
All these data were well described by a global 
fit~\cite{Anselmino:2007fs,Anselmino:2008jk} which allowed for a first 
extraction of the $u$ and $d$-quark transversity PDFs.

The Sivers asymmetry for the proton was measured by 
HERMES~\cite{Airapetian:2004tw,Airapetian:2009ti}
to be different from zero for positive hadrons, while
it was found to be compatible with zero for deuteron
by COMPASS ~\cite{Alexakhin:2005iw,Ageev:2006da,Alekseev:2008dn}.
These HERMES and COMPASS data could also be well described by theoretical 
calculations and fits, and  allowed for extractions of the Sivers function
\cite{Anselmino:2005an}, which turned out to be different from zero and 
opposite in sign for $u$ and $d$-quarks.

In this Letter, we present the  COMPASS results on the Collins and Sivers 
asymmetries for charged hadrons produced in SIDIS of high energy muons on  
transversely polarised protons.
The data were collected in 2007 using NH$_3$ as target material
and a 160~\gevc\ beam with a momentum spread  $\Delta p / p = \pm 5$\%.
The beam was naturally polarised by the $\pi$--decay mechanism, with a
longitudinal polarisation of about -80\%.
This measurement followed the  measurements performed in 2002, 
2003 and 2004 at the same energy with the transversely polarised 
$^6$LiD target.

The COMPASS spectrometer~\cite{Abbon:2007pq} is in operation on the 
M2 beam line of CERN since 2002.
Two magnetic stages are used to ensure large angular and momentum acceptance.
A variety of tracking detectors is used to cope with the
different requirements of position accuracy and rate capability at different 
angles.
Particle identification is provided by a large acceptance RICH detector, 
calorimeters, and muon filters.
Major upgrades in 2005 mainly concerned the polarised target, the tracking 
system, the RICH detector, and the electromagnetic calorimeters.
The new  target solenoid magnet provides a field of 2.5~T
and has a polar angle acceptance of 180~mrad as seen from the 
upstream end of the target.
In the earlier measurements with the $^6$LiD target the polar angle acceptance
was  70~mrad.
The target material is cooled in a $^3$He--$^4$He dilution
refrigerator, and the protons in the H atoms are polarised to 0.80--0.90 by
dynamical nuclear polarisation.
About 48 hours are necessary to reach 95\% of the maximal polarisation.
A pair of saddle-shaped coils  can provide a 0.6~T vertical field
which is used  to rotate the target nucleon spin and to hold
the polarisation vertical for the transversity measurements.
In the frozen spin mode, and with the holding field at its operational 
value, the relaxation time of the polarisation exceeds 3000 hours.

The  target consisted of three cylindrical cells with 4~cm diameter,
one central cell of 60~cm length and two outer ones of 30~cm length, 
all separated by 5~cm.
Neighbouring cells were polarised in opposite directions, so that data 
with both spin directions were recorded at the same time.
In order to minimise the effects due to  different spectrometer acceptance 
for  different target cells, in each period of data taking a polarisation 
reversal was performed after 4--5 days by changing the microwave 
frequencies in the three cells.

The geometry of the polarised target and the data taking procedure
were chosen such as to optimise the extraction of spin asymmetries. 
The principle of the measurement can be understood by considering 
the ``ratio product''~\cite{Ageev:2006da}
\be
R =
\frac{N^{\uparrow}_{inner}}{N^{\downarrow}_{inner}} \cdot 
\frac{N^{\uparrow}_{outer}}{N^{\downarrow}_{outer}}  \, ,
\label{eq:rp}
\ee
where $N^{\uparrow}_{inner}$ and $N^{\downarrow}_{outer}$ are
the number of hadrons produced in the first sub-period on oppositely 
polarised cells, and $N^{\downarrow}_{inner}$ and $N^{\uparrow}_{outer}$ 
are the corresponding numbers in the second sub-period, i.e. after 
polarisation reversal.
The ratio product is constructed such that beam flux, spin-averaged 
cross-section, and the number of scattering centres cancel.
As long as the ratios between the spectrometer acceptances
of each cell are the same in the two sub-periods and the number of 
produced hadrons follows the generic azimuthal modulation 
$N^{\uparrow \downarrow}\sim 1 \pm \epsilon \sin \Phi$, one simply gets
$R =1+4\epsilon \sin \Phi$, and the extraction of the amplitude 
$\epsilon$ of the azimuthal modulation is straightforward.

In 2007 data were taken at a  mean beam intensity of about
$5 \times 10^7 \; \mu^+$/s (typically $2.4 \times 10^8 \; \mu^+$/spill,
for a spill length of 4.8 s every 16.8 s).
Using up $4 \times 10^{13}$ muons, about $12 \times 10^9$ events were 
collected in six separate periods, corresponding to 440~TB of data.

In the data analysis, events  were selected if they had 
at least one ``primary vertex'', defined as the intersection point 
of a beam track, the scattered muon track, and other possible outgoing 
tracks. 
The momenta of both incoming and outgoing charged particles were measured.
The primary vertex was required to be inside a target cell. 
In order to guarantee the same muon flux  along the target  material,
the extrapolated beam track  had to traverse all the three target cells.
For  incoming and  scattered muon tracks, as well as for
the other reconstructed tracks, $\chi^2$ cuts were applied
to assure the quality of track reconstruction.
Tracks from the primary vertex which  traversed more than
30 radiation lengths were identified as scattered muons.
The event was rejected if more than one of such tracks were found.

In order to be in the DIS regime, only events with a photon virtuality 
$Q^2>1$~(\gevc)$^2$, a fractional energy of the virtual photon 
$0.1<y<0.9$, and a mass of the hadronic final state $W>5$~\gevcc\ were 
considered.
The variable $x$ covers the range from 0.004 to 0.7.

All particles emerging from the primary vertex were assumed to be hadrons
if they traversed less than 10 radiation lengths of material. 
For tracks with an associated  cluster  in one of the hadronic 
calorimeters, a minimal amount of deposited energy
was required to further reduce the electron and muon contamination.
Finally, tracks reconstructed only in the fringe field of 
the first analysing magnet of the spectrometer were rejected. 
This roughly corresponds to a cut at 1.5~\gevc\ in the 
hadron momenta. 
In order to reconstruct the hadron azimuthal angle with good precision,
the hadron transverse momentum with respect to the virtual photon direction,
$p_T^h$, was required to be above 0.1~\gevc .
A minimum value of 0.2 for $z$, the relative energy of the hadron with
respect to the virtual photon energy, was chosen to  avoid hadrons from
the target fragmentation region.

As explained in detail in Ref.~\cite{Ageev:2006da},
the Collins effect shows up as a modulation 
$[1+\epsilon_C \sin (\phi_h+\phi_S-\pi)]$ in the number of events,
where $\phi_h$ and $\phi_S$ are the azimuthal angles of the hadron and 
of the target nucleon spin vector in a reference system in which the 
z-axis is the virtual photon direction and the x--z plane is the 
lepton plane according to Ref.~\cite{Bacchetta:2004jz}.
.
The amplitude of the modulation is  
$\epsilon_C=D_{NN} f P_T A_{Coll}$, where $D_{NN}= (1-y)/(1-y + y^2/2)$ 
is the transverse spin transfer coefficient from  target quark 
to  struck quark, $f$ the dilution
factor of the NH$_3$ material, and 
$P_T$ is the proton polarisation.
Similarly, the Sivers effect results in a modulation
 $[1+\epsilon_S \sin (\phi_h-\phi_S)]$, where
$\epsilon_S= f P_T A_{Siv}$.

The transverse spin asymmetries were  obtained by comparing
the azimuthal distributions of the detected hadrons as measured in 
the first sub-period of data taking with the corresponding distributions 
of the second half measured with opposite target polarisation.
Since the two sets of data were taken typically one week apart, the 
stability of the apparatus is a central point in the measurement.
As a first step in the data selection, the hit distributions of all trackers 
were scrutinised, as well as the number
of reconstructed events, the number of vertices per events, and the number
of tracks per event.
In a second step, the stability of the average $\pi^+ \pi^-$ 
invariant mass in the $K^0$ region as well as the distribution
of twelve kinematic quantities ($x, \, y, \, W, \, z, \,  ...$) 
were investigated dividing the data in small time-ordered sub-samples.
Each distribution of each sub-sample was compared with the corresponding 
ones of each other sub-sample within the same data taking period, and 
sub-samples were rejected when deviating more than 3.5~$\sigma_{stat}$
from the mean values.

As a final selection criterion, the data were tested for
a possible dependence  on either
$\sin (\phi_h+\phi_S)$ or $\sin (\phi_h-\phi_S)$ of the
acceptance ratio between two consecutive sub-periods with opposite
target polarisation.
Combining the number of events reconstructed in the different target 
cells in two consecutive data taking sub-periods, one can construct 
two different estimators on the stability of the acceptance.
The first estimator measures the mean modulation in the relevant 
azimuthal angle of the acceptance ratio between two sub-periods.
The second one probes possible large differences in the acceptance ratios for
the different target cells which could affect the physics asymmetry.
These two pieces of information have been used to construct a $\chi^2$ and 
the final selection of the data taking periods was done on the basis 
of its  value.

As a result of the quality control, all data collected in the six periods
were used for the extraction of the Collins asymmetry, while only
four periods were used for the Sivers asymmetry.
This can be understood because the Sivers asymmetry is very 
sensitive to  instabilities of the spectrometer since it is
the amplitude of a modulation of the azimuthal  angle of the hadron
transverse momentum with respect to the target spin vector.
On the contrary, the Collins asymmetry is an asymmetry in the azimuthal angle
between the hadron transverse momentum and a direction which depends on 
the target spin direction and the lepton scattering plane,
which is different for each event.
The final sample contains $23.1 \times 10^6$ SIDIS events for 
the Collins asymmetry and $15.6 \times 10^6$ for the Sivers asymmetry.

The asymmetries were evaluated for positive and negative hadrons 
in bins of the three kinematic variables $x$, $z$ and $p_T^h$.
The binning is the same as used for the previous analyses of deuteron data
and consists of 9 bins in $x$,   8 bins in $z$ and  9 bins in $p_T^h$,
integrating over the other two variables.
For each period, the physics asymmetries were obtained
by dividing the raw asymmetries by the target polarisation,
the dilution factor, and, in the case of Collins analysis,
by the $D_{NN}$ factor.
The target polarisation was measured individually for each cell and each 
period.
The dilution factor of the ammonia target was evaluated for each bin.
It is 0.15 in average, and increases with $x$ from 0.14 to 0.17.

The estimator used for the evaluation of the raw asymmetries is based on an 
extended unbinned maximum likelihood method~\cite{Barlow:1990vc}.
The likelihood function is built as the  product of the probability 
densities  $p$ corresponding to each hadron $i$ from each target cell.
The likelihood for hadrons from a given target cell in one period 
is written as
\be
\mathcal{L}  = \left(e^{-I^+}\prod_{i=0}^{N^+}p^+(\phi_{h,i},\phi_{S,i})
\right)^{\frac{1}{N^+}}\cdot 
\left(e^{-I^-}\prod_{i=0}^{N^-}p^-(\phi_{h,i},\phi_{S,i})\right)^{\frac{1}{N^-}}\, .
\ee
The $+$ and $-$ signs refer to the orientation of the target polarisation
in the two sub-periods and $N^{\pm}$ is the corresponding total
number of hadrons.
The quantities $I^{\pm}$ are the integrals of the probability densities 
over $\phi_S$ and $\phi_h$.
The probability densities $p^{\pm}$ are the product of two parts, 
one corresponding to the acceptance description and the other 
to the SIDIS cross section of longitudinally polarised leptons on 
transversely polarised nucleons.
Various parametrisations of the acceptance part were tested,
resulting in a negligible dependence of the extracted asymmetries 
on the acceptance description.
The  cross section was parametrised taking into account both the unpolarised 
and polarised parts.
The polarised part consists of all the expected eight modulations,
namely
$\sin (\phi_h+\phi_S-\pi)$, $\sin (\phi_h-\phi_S)$, $\cos (\phi_h-\phi_S)$, 
$\sin (2\phi_h-\phi_S)$, $\cos (2\phi_h-\phi_S)$, 
$\sin (\phi_S)$, $\cos (\phi_S)$, and
$\sin (3\phi_h-\phi_S)$,
and all their amplitudes were extracted at the same time.
The Collins and Sivers asymmetries are proportional to the amplitudes of 
the first two terms.
Systematic studies for the other six amplitudes are still ongoing, and 
those results will be the subject of a future publication.

The final Collins and Sivers asymmetries extracted with the likelihood 
method were compared  with the asymmetries extracted
using four other estimators, including those used in the
previous publications which were based on the ``ratio 
product'' $R$ of Eq.~\ref{eq:rp},
finding an excellent agreement between all  results. 
The correlation coefficient between
the Collins and the Sivers asymmetries turned out to 
be small, less than 0.2 in absolute value over the whole $x$ range.

Extensive  studies were performed in order to assess the 
systematic uncertainty of the measured asymmetries. 
All the studies were done separately for positive and negative hadrons 
and for the Collins and the Sivers asymmetries.

The largest systematic error is due to  residual acceptance
variations within  pairs of data taking sub-periods.
To quantify these effects,
two different types of false asymmetries were calculated,
using the  external cells and the internal cell divided in two parts, 
and assuming wrong sign polarisation for one of the two. 
Moreover, the physical asymmetries were also extracted 
using only the first and only the second half of the target.
The difference between these two physical asymmetries,
the false asymmetries, and the degree of compatibility of the results 
from different periods were all used to quantify the systematic uncertainty. 

In the case of the Collins asymmetry, the systematic uncertainty
is estimated to be 0.5~$\sigma_{stat}$ for positive and 
0.6~$\sigma_{stat}$ for negative hadrons.
In the case of the Sivers asymmetry, the systematic error
is 0.8~$\sigma_{stat}$ for positive and 0.4~$\sigma_{stat}$ for negative 
hadrons.
A further systematic uncertainty of $\pm 0.01$ is present in the absolute
scale of the Sivers asymmetry for positive hadrons.
It reflects a 0.02 difference in the mean value of the asymmetries
extracted in the first two and in the second two periods 
of data taking used for this analysis.
In spite of throughout studies, the origin of this difference, which 
affects only the Sivers asymmetry for positive hadrons, could not be
identified and had therefore to be included in the systematic
uncertainty.
\begin{figure}[tb]
 \includegraphics[width=0.99\textwidth]{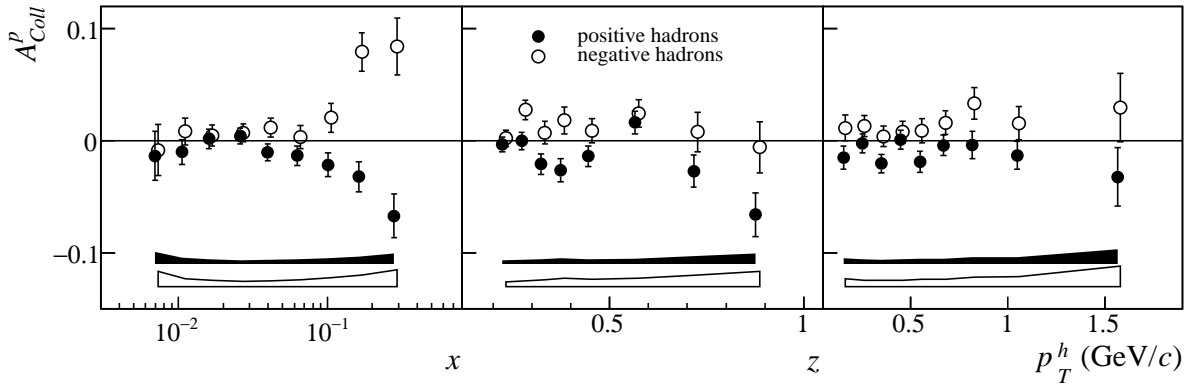}
 \caption{Collins asymmetry
 as a function of $x$, $z$, and $p_T^h$, for positive (closed points) 
 and negative (open points) hadrons. 
 The bars show the statistical errors.
 The point to point systematic uncertainties have been estimated to be
0.5~$\sigma_{stat}$ for positive and 
0.6~$\sigma_{stat}$ for negative hadrons
and are given by the bands. }
\label{fig:res_c}  
\end{figure}
\begin{figure}[tbh]
\begin{center}
 \includegraphics[width=0.99\textwidth]{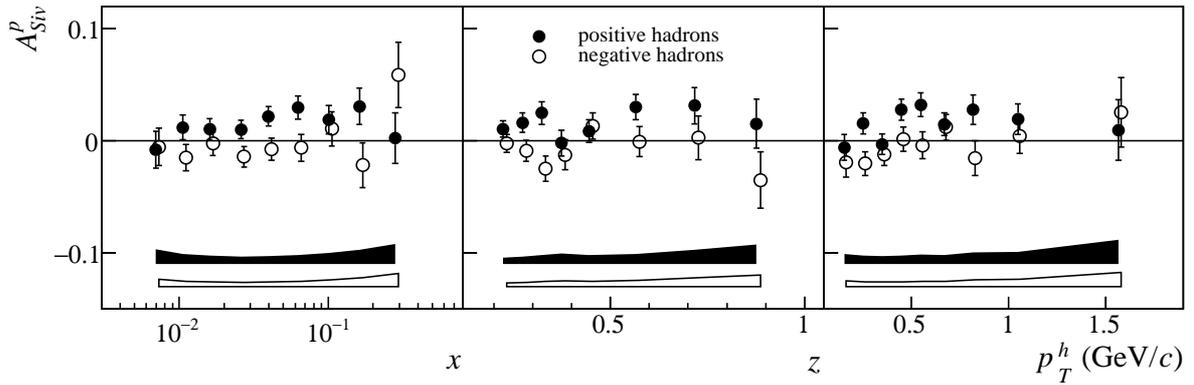}
 \caption{Sivers asymmetry
as a function of $x$, $z$, and $p_T^h$, for positive (closed points) 
and negative (open points) hadrons. 
The bars show the statistical errors.
The point to point systematic uncertainties
have been estimated to be
0.8~$\sigma_{stat}$ for positive and 
0.4~$\sigma_{stat}$ for negative hadrons
and are given by the bands. 
For positive hadrons only, an absolute scale uncertainty of
$\pm 0.01$ has also to be taken into account.}
\label{fig:res_s}  
\end{center}
\end{figure}
\begin{figure}[tbh!]
\begin{center}
 \includegraphics[width=0.99\textwidth]{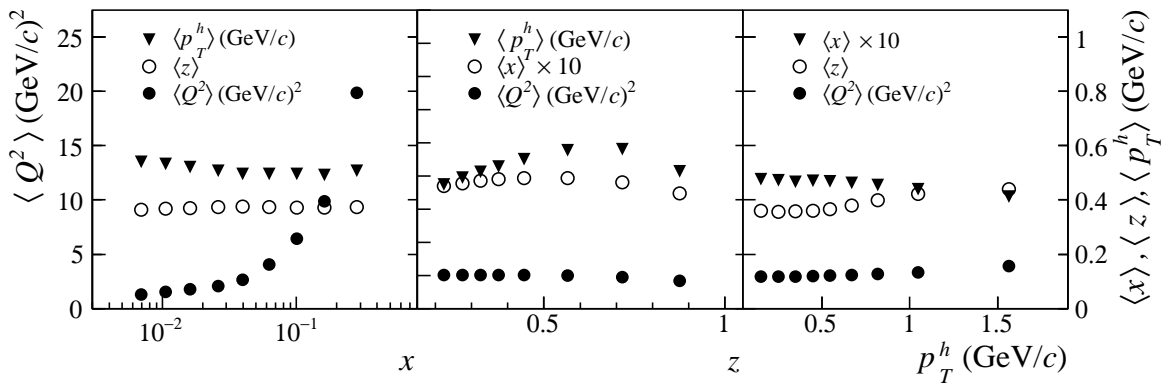}
 \caption{Mean values of some kinematic variables in the
final data sample.
From left to right: mean values of $p_T^h$, $z$ and $Q^2$
as functions of $x$; mean values of $p_T^h$, $x$ and $Q^2$
as functions of $z$; mean values of $x$, $z$ and $Q^2$
as functions of $p_T^h$.}
\label{fig:res_kin}  
\end{center}
\end{figure}
The results of this measurement of the Collins and Sivers asymmetries 
are shown in Fig.~\ref{fig:res_c} and~\ref{fig:res_s} as a function of 
$x$, $z$, and $p_T^h$, for positive and negative hadrons.
Figure~\ref{fig:res_kin} displays the mean values of kinematic variables
for positive hadrons 
in the $x$, $z$, and $p_T^h$ bins.
The corresponding quantities for negative hadrons are very
similar\footnote{
All  numerical values have been put to HEPDATA.
}.

As it is clear from Fig.~\ref{fig:res_c}, the Collins asymmetry 
has a strong $x$ dependence. It is  compatible with zero
at small $x$ within the small statistical errors and increases in
absolute value up to about 0.1 for $x>0.1$.
There, the values agree both in magnitude and in sign
with the previous measurements of HERMES~\cite{Airapetian:2004tw},
which were performed at the considerably lower electron beam 
energy of 27.5~GeV.
Also, the present results agree with the predictions of the global 
analysis of ref.~\cite{Anselmino:2007fs,Anselmino:2008jk} and thus
strongly support the underlying interpretation
of the Collins asymmetry in terms of a convolution of the twist-two 
transversity PDF and  the FF of a transversely polarised quark.
An important issue  is the $Q^2$ dependence of these functions.
Our results at large $x$ are compatible with the HERMES data in spite of the
higher $Q^2$ values which exceed those of HERMES
by a factor 2 to 3 with increasing $x$.
This indicates that the possible $Q^2$ dependence should not
be dramatic in the present energy ranges.
\begin{figure}[tb]
\begin{center}
 \includegraphics[width=0.8\textwidth]{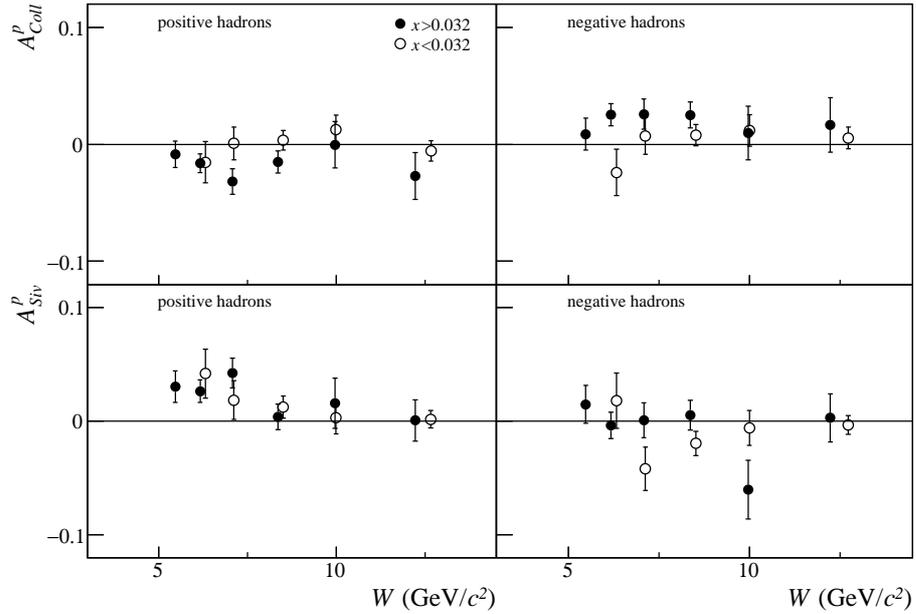}
 \caption{Collins (upper row) and Sivers (lower row) asymmetry 
as a function of $W$, for positive (left) 
and negative (right) hadrons. 
The closed and open points give the values for the ``large $x$''
and the ``small $x$'' samples respectively.
The errors are statistical only.
}
\label{fig:res_w}  
\end{center}
\end{figure}
\begin{figure}[tbh!]
\begin{center}
\includegraphics[width=0.8\textwidth]{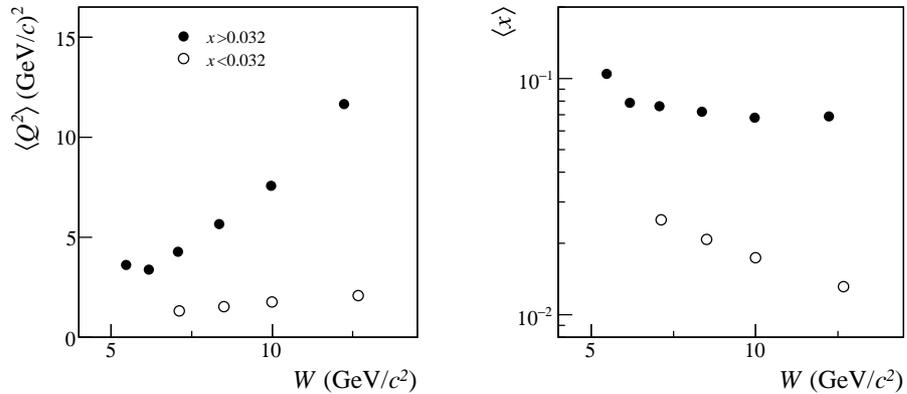}
 \caption{Mean values of $Q^2$ (left) and $x$ (right)
as functions of $W$.
The closed and open points give the values for the ``large $x$''
and the ``small $x$'' samples respectively.}
\label{fig:res_wk}  
\end{center}
\end{figure}

The results for the Sivers asymmetry for 
negative hadrons exhibit values compatible with zero within the
statistical accuracy of the measurement.
For positive hadrons, the data indicate small positive values,
up to about 3\% in the valence region.
These values are somewhat smaller  than but still
compatible with the ones measured by HERMES at smaller $Q^2$.
Given the importance of the Sivers function in the present description
of the transverse momentum structure of the nucleon,
we  looked at a possible kinematic dependence of our measurements.
In particular, we  evaluated the asymmetries
as a function of $W$, separately for the ``large-$x$'' ($x>0.032$) and 
``small-$x$'' ($x<0.032$) samples.
The results are shown in Fig.~\ref{fig:res_w}.
The mean values of $Q^2$ and $x$ in all  $W$ bins are given
in Fig.~\ref{fig:res_wk}.
As it is apparent from Fig.~\ref{fig:res_w}, no conclusion can be drawn
about a possible $W$ dependence of the Collins asymmetry.
On the other hand, the signal of the Sivers asymmetry for
positive hadrons seems to be  concentrated at small $W$,
in the region where HERMES measures, and
goes to zero at large $W$, which for large $x$ means large $Q^2$.
Thus our data give an indication for a possible  $W$
dependence of the Sivers asymmetry for positive hadrons. 
Definite conclusions will be possible only when new more precise
data at high energy will become available.

In summary, for the first time the  Collins and  Sivers asymmetries
for positive and negative hadron production in DIS off the proton have 
been measured at high energy.
Our data extend the kinematic range to large $Q^2$ and large $W$ values.
The $x$ range has been extended to considerably smaller values 
which are needed to evaluate the PDF first moments.
For the Sivers asymmetry, a signal is seen for positive hadrons,
which persists to rather small $x$ values.
The data give an indication for a possible $W$ dependence
of this asymmetry, but the present statistical  and systematic 
uncertainties do not allow definite conclusions.
The measured Collins asymmetry is sizable  for both positive 
and negative hadrons also at high energies and $Q^2$.
Thus Collins asymmetries measured in SIDIS are an appropriate tool to 
investigate the transversity PDF.

\section*{Acknowledgements}
We gratefully acknowledge the
support of the CERN management and staff and the skill and effort
of the technicians of our collaborating institutes. Special
thanks go to V.~Anosov and V.~Pesaro for their technical support
during the installation and running of this experiment.

\end{document}